\documentclass[a4paper,12pt]{article}
\usepackage{amsmath}
\usepackage[psamsfonts]{amssymb}
\usepackage{rsfs}
\usepackage{bm}
\usepackage{eucal}

\usepackage{cite}
\usepackage{graphicx}
\usepackage{tabularx}
\usepackage{layout}
\usepackage{color}

\begin{document} 

\begin{titlepage}

\hrule 
\leftline{}
\leftline{Chiba Univ. Preprint
          \hfill   \hbox{\bf CHIBA-EP-125}}
\leftline{\hfill   \hbox{hep-th/0103141}}
\leftline{\hfill   \hbox{March 2001}}
\vskip 5pt
\hrule 
\vskip 1.0cm
\centerline{\large\bf 
 Spontaneous breakdown of 
} 
\vskip 0.5cm
\centerline{\large\bf  
BRST supersymmetry
}
\vskip 0.5cm
\centerline{\large\bf  
due to ghost condensation in QCD
}

\vskip 1cm

\centerline{{\bf 
Kei-Ichi Kondo$^{1,}{}^{\dagger}{}$
}}  
\vskip 1cm
\begin{description}
\item[]{\it \centerline{ 
$^1$Department of Physics, Faculty of Science, 
Chiba University,  Chiba 263-8522, Japan}
  }
\end{description}

\centerline{{\bf Abstract}} 
We argue that the BRST and the anti-BRST super symmetries in the four-dimensional Yang-Mills theory can be spontaneously broken in a nonlinear partial gauge due to ghost--anti-ghost condensation. 
However, we show that the spontaneous BRST symmetry breaking can be avoided if we adopt the modified Maximal Abelian gauge which is an orthosymplectic $OSp(4|2)$ invariant renormalizable gauge proposed by the author to derive quark confinement.  
We compare the Maximal Abelian gauge with the conventional $OSp(4|2)$ invariant gauge proposed by Delbourgo--Jarvis and Baulieu--Thierry-Mieg.
Finally, an implication to the Gribov problem is briefly mentioned.


\vskip 0.2cm
$^\dagger$ 
  E-mail:  {\tt kondo@cuphd.nd.chiba-u.ac.jp}
\vskip 0.2cm  
\hrule  

\par 
\par\noindent


\vskip 0.5cm  


\pagenumbering{roman}
\tableofcontents
\pagenumbering{arabic}

\vskip 0.5cm  
\hrule  

\end{titlepage}


\section{\label{sec:intro}Introduction}
\setcounter{equation}{0}

In modern particle physics, all the fundamental forces are considered to be mediated by the corresponding gauge bosons.  The gauge boson is described by the gauge theory which is characterized by its invariance under the gauge transformation of the fundamental fields.   The gauge invariance is a guiding principle to determine the form of the interaction between gauge fields and matter fields through the minimal coupling.
In quantizing the gauge theory, however, we need to fix (or reduce) the gauge degrees of freedom, i.e., the redundant degrees of freedom associated with the gauge transformation.
After the prescription of gauge fixing, the gauge theory no longer has the gauge symmetry.
Nevertheless, it has been discovered that the gauge theory after the gauge fixing possesses a kind of global supersymmetry called the Becchi-Rouet-Stora-Tyutin (BRST) symmetry~\cite{BRST} which plays the fundamental role in the gauge theory. In fact, all the Ward-Takahashi and Slavnov-Taylor identities associated with the local gauge symmetry are derived from the global BRST supersymmetry.
\par
 The BRST symmetry is a continuous global symmetry, while the gauge symmetry is a continuous local symmetry. 
For the Yang-Mills gauge theory, the BRST transformation is given by
\begin{subequations}
\begin{equation}
\begin{align}
 \bm{\delta}_B \mathscr{A}_\mu(x) &=
\mathscr{D}_\mu[\mathscr{A}]\mathscr{C}(x) :=
 \partial_\mu \mathscr{C}(x) + g (\mathscr{A}_\mu(x) \times \mathscr{C}(x)) , \\
 \bm{\delta}_B \mathscr{C}(x) &= -{1 \over 2}g(\mathscr{C}(x) \times \mathscr{C}(x)) ,
\\
 \bm{\delta}_B \bar{\mathscr{C}}(x) &= i \mathscr{B}(x) ,
\\
 \bm{\delta}_B \mathscr{B}(x) &= 0 ,
\end{align}
\label{BRST1}
\end{equation}
\end{subequations}
where%
\footnote{We use the following notation:
\begin{equation}
 F \cdot G := F^A G^A, \quad 
 F^2 := F \cdot F , \quad 
 (F \times G)^A := f^{ABC}F^B G^C ,
\nonumber
\end{equation}
where $f^{ABC}$ are the structure constants of the Lie algebra $\mathscr{G}$ of the gauge group $G$.
}
$\mathscr{A}_\mu, \mathscr{B}, \mathscr{C}$ and $\bar{\mathscr{C}}$ are the gauge field, Nakanishi-Lautrap (NL) auxiliary field, Faddeev--Popov (FP) ghost and anti-ghost fields respectively.
It is important to notice that the ghost field $\mathscr{C}$ and the anti-ghost field $\bar{\mathscr{C}}$ are Hermitian fields which are mutually independent, i.e.,
$\mathscr{C}^\dagger(x) = \mathscr{C}(x)$,
$\bar{\mathscr{C}}^\dagger(x) = \bar{\mathscr{C}}(x)$.
(Note that $\bar{\mathscr{C}}$ is neither the complex conjugate $\mathscr{C}^*$ of $\mathscr{C}$ nor the Hermitian conjugate $\mathscr{C}^\dagger$ of $\mathscr{C}$). 
By this assignment, the total Lagrangian of the Yang-Mills theory with the gauge fixing (GF) and FP ghost term becomes hermitian.
As suggested from the observation that the role of $\mathscr{C}$ and $\bar{\mathscr{C}}$ is asymmetric in the BRST transformation,  
there is another BRST transformation, say anti-BRST transformation~\cite{antiBRST}), given by
\begin{subequations}
\begin{equation}
\begin{align}
 \bar{\bm{\delta}}_B \mathscr{A}_\mu(x) &= 
\mathscr{D}_\mu[\mathscr{A}] \bar{\mathscr{C}}(x) :=
\partial_\mu \bar{\mathscr{C}}(x) + g (\mathscr{A}_\mu(x) \times \bar{\mathscr{C}}(x)) , \\
 \bar{\bm{\delta}}_B \bar{\mathscr{C}}(x) &= -{1 \over 2}g(\bar{\mathscr{C}}(x) \times \bar{\mathscr{C}}(x)) ,
 \\
 \bar{\bm{\delta}}_B \mathscr{C}(x) &= i \bar{\mathscr{B}}(x) ,
\\
 \bar{\bm{\delta}}_B \bar{\mathscr{B}}(x) &= 0 ,
\end{align}
\label{BRST2}
\end{equation}
\end{subequations}
where $\bar{\mathscr{B}}$ is defined by
\begin{equation}
  \bar{\mathscr{B}}(x) = -\mathscr{B}(x) + ig (\mathscr{C}(x) \times \bar{\mathscr{C}}(x)) .
\end{equation}
\par
It is well known that the manifest covariant formulation of the gauge theory needs an indefinite inner product space $\mathscr{V}$.  The BRST symmetry plays the crucial role to single out the physical state space $\mathscr{V}_{\text{phys}}$ with positive semi-definite metric as a subspace of $\mathscr{V}$.  In fact, it has been shown by Kugo and Ojima~\cite{KO79} that the physical state is invariant under the BRST transformation, in other words, the physical state is characterized by the generator $Q_B$ of the BRST transformation (i.e., the BRST charge $Q_B$) as
\begin{equation}
  Q_B | {\rm phys} \rangle = 0 ,
\label{psc1}
\end{equation}
where $\bm{\delta}_B {\cal O} = [ iQ_B, {\cal O} ]_{\pm}$.
Similarly to (\ref{psc1}), the following condition is also imposed:
\begin{equation}
  \bar{Q}_B | {\rm phys} \rangle = 0 ,
\label{psc2}
\end{equation}
without any inconsistency and without any change in the physical contents of the theory.
Moreover, an additional condition for the ghost number operator $Q_c$ can be imposed: 
\begin{equation}
  Q_C | {\rm phys} \rangle = 0 .
\label{psc3}
\end{equation}
The generators $Q_B, \bar{Q}_B, Q_c$ form the BRST algebra~\cite{antiBRST}:
\begin{equation}
\begin{align}
 i [ Q_B , Q_C ] &= - Q_B,  & i[ \bar{Q}_B,Q_C ] &= \bar{Q}_B , & {} 
 \nonumber\\
 \{ Q_B, Q_B \} &= 0,  &  \{ \bar{Q}_B, \bar{Q}_B \} &= 0, &
   \{ Q_B, \bar{Q}_B \} = 0 .
\end{align}
\end{equation}
This relation implies that $Q_B (\bar{Q}_B)$ generates the translation $\theta \rightarrow \theta + \lambda 
(\bar{\theta} \rightarrow \bar{\theta}+\bar{\lambda})$ and $Q_c$ the dilatation 
$\theta \rightarrow e^{\rho}\theta$ with $\theta$ and $\lambda$ being the real elements of the Grassmann algebra and $\rho$ a real number \cite{BT81}. The BRST symmetry is a supersymmetry in the sense that the BRST transformation changes the commuting variable into the anti-commuting one and vice versa.  
\par
In the paper~\cite{Fujikawa83}, however, Fujikawa emphasized that the BRST supersymmetry in non-Abelian gauge theories is not what it to be imposed on the theory but rather what is to be proved.  The spontaneous breakdown of the BRST symmetry is demonstrated by a non-gauge model and it is examined also in non-Abelian gauge theories.  It is pointed out that the dynamical stability of BRST supersymmetry is closely related to the so-called Gribov problem for the Lorentz type gauge 
$\partial_\mu \mathscr{A}^\mu = 0$.
\par
In this paper, we discuss the spontaneous BRST supersymmetry breaking in the Yang-Mills theory with the  GF plus  FP ghost term $\mathscr{L}_{GF+FP}$ which has the  characteristic features enumerated below.
In particular, we discuss the Delbourgo--Jarvis--Baulieu--Thierry-Mieg (DJBT) gauge~\cite{DJ82,BT82} and the modified Maximally Abelian (MA) gauge~\cite{KondoII}.  Both gauge satisfy the following properties:
\begin{enumerate}
\item
 $\mathscr{L}_{GF+FP}$ is BRST invariant, i.e., $\bm{\delta}_B \mathscr{L}_{GF+FP}=0$, due to nilpotency of the BRST transformation $\bm{\delta}_B^2 \equiv 0$.

\item
 $\mathscr{L}_{GF+FP}$ is anti-BRST invariant, i.e., $\bar{\bm{\delta}}_B \mathscr{L}_{GF+FP}=0$, due to nilpotency of the anti-BRST transformation $\bar{\bm{\delta}}_B^2 \equiv 0$.

\item
 $\mathscr{L}_{GF+FP}$ is invariant under the $OSp(4|2)$ rotation among the component fields in the supermultiplet $(\mathscr{A}_\mu, \mathscr{C}, \bar{\mathscr{C}})$ defined on the superspace $(x_\mu, \theta, \bar{\theta})$.  The hidden supersymmetry causes the dimensional reduction in the sense of Parisi-Sourlas.  Then the 4-dimensional GF+FP sector reduces to the 2-dimensional nonlinear sigma model.  See ref.~\cite{HK85,KondoII} for more details.

\item
 $\mathscr{L}_{GF+FP}$ is invariant under the FP ghost conjugation,
\begin{equation}
 \mathscr{C}^A \rightarrow \pm \bar{\mathscr{C}}^A, \quad
 \bar{\mathscr{C}}^A \rightarrow \mp \mathscr{C}^A, \quad
 \mathscr{B}^A \rightarrow - \bar{\mathscr{B}}^A, \quad
 \bar{\mathscr{B}}^A \rightarrow - \mathscr{B}^A, \quad
 (\mathscr{A}_\mu^A \rightarrow \mathscr{A}_\mu^A) .
\end{equation}
Therefore, $\mathscr{C}$ and $\bar{\mathscr{C}}$ can be treated on equal footing.  In other words, the theory is totally symmetric under the exchange of $\mathscr{C}$ and $\bar{\mathscr{C}}$.

\item
The total Yang-Mills theory with the GF+FP term $\mathscr{L}_{GF+FP}$ 
is (multiplicatively) renormalizable, see~\cite{DJ82,Baulieu85} and~\cite{MLP85}.
In particular, the naive MA gauge 
\begin{equation}
  \mathscr{L}_{GF+FP} = - i \bm{\delta}_B \left[ 
\bar{C}^a  \left\{ D_\mu[a]A^\mu + {\alpha \over 2} B \right\}^a \right] 
\end{equation}
spoils the renormalizability.  This is because the MA gauge is a nonlinear gauge.  For the renormalizability of the Yang-Mills theory in the MA gauge, therefore, we need the four-ghost interaction from the beginning.  In fact, the renormalizability of the Yang-Mills theory in Abelian gauge supplemented with the four-ghost interaction was proved to all orders in perturbation theory~\cite{MLP85}.

\end{enumerate}
In the following argument, thus, existence of both the BRST symmetry and the anti-BRST symmetry is essential.
In this paper we discuss the possibility of the spontaneous breakdown of the BRST or anti-BRST symmetry due to {\it ghost--anti-ghost condensation}.%
\footnote{
In this paper, we assume that the ghost number is not spontaneously broken, i.e.,  
$Q_C | {\rm phys} \rangle = 0$,
even if 
$Q_B | {\rm phys} \rangle \not= 0$, or
$\bar{Q}_B | {\rm phys} \rangle \not= 0$.
Therefore, we have for example
$\langle C \times C \rangle = 0 = \langle \bar{C} \times \bar{C} \rangle$.
}
  This type of condensation has been proposed recently~\cite{Schaden99,KS00} as a mechanism of providing the masses of off-diagonal gluons and off-diagonal ghosts in the Yang-Mills theory in the Maximally Abelian gauge.  This mechanism can give an evidence of the infrared Abelian dominance~\cite{tHooft81}, thereby justifies the dual superconductor picture~\cite{dualsuper} of QCD vacuum for explaining quark confinement~\cite{KondoI,KondoII,KondoIII,KondoIV,KondoV,KondoVI,Kondo00}.  Finally, we discuss a relationship between the spontaneous breaking of BRST symmetry and the Gribov problem~\cite{Gribov78}.

\par
\section{\label{sec:DJgauge}Delbourgo--Jarvis and Baulieu--Thierry-Mieg  gauge}

First, we examine an orthosymplectic $OSp(4|1)$ invariant gauge fixing term proposed by Delbourgo--Jarvis and Baulieu--Thierry-Mieg (DJBT) ~\cite{DJ82,BT82}.  For the gauge group $G=SU(N)$, it reads

\begin{equation}
 \mathscr{L}_{GF+FP} = i \bm{\delta}_B \bar{\bm{\delta}}_B \left( {1 \over 2}\mathscr{A}_\mu^A \mathscr{A}^\mu{}^A - {\alpha \over 2}i \mathscr{C}^A \bar{\mathscr{C}}^A \right) 
\quad (A=1, \cdots, N^2-1) .
\end{equation}
By performing the BRST transformation explicitly, the GF+FP term is rewritten as
\begin{equation}
\begin{align}
 \mathscr{L}_{GF+FP} &= \mathscr{B} \cdot \partial_\mu \mathscr{A}^\mu + i \bar{\mathscr{C}} \cdot \partial_\mu \mathscr{D}^\mu[\mathscr{A}]\mathscr{C}
\nonumber\\&
+ {\alpha \over 2} \mathscr{B} \cdot \mathscr{B} - {\alpha \over 2} ig (\mathscr{C} \times \bar{\mathscr{C}}) \cdot \mathscr{B} 
+ {\alpha \over 8}g^2 (\bar{\mathscr{C}} \times \bar{\mathscr{C}}) \cdot (\mathscr{C} \times \mathscr{C}) 
\\
&= \mathscr{B} \cdot \partial_\mu \mathscr{A}^\mu + i \bar{\mathscr{C}} \cdot \partial_\mu \mathscr{D}^\mu[\mathscr{A}]\mathscr{C}
\nonumber\\&
+ {\alpha \over 2}\mathscr{B} \cdot \mathscr{B} - {\alpha \over 2} ig (\mathscr{C} \times \bar{\mathscr{C}}) \cdot \mathscr{B} 
+ {\alpha \over 4}g^2 (i \mathscr{C} \times \bar{\mathscr{C}}) \cdot (i \mathscr{C} \times \bar{\mathscr{C}}) .
\end{align}
\end{equation}

\par
First, we examine the case of non-zero $\alpha$. 
The GF+FP term can be recast into
\begin{equation}
\begin{align}
 \mathscr{L}_{GF+FP} 
&= 
{\alpha \over 2}\left( \mathscr{B} - {1 \over 2}ig (\mathscr{C} \times \bar{\mathscr{C}}) + {1 \over \alpha} \partial_\mu \mathscr{A}^\mu \right)^2 
- {1 \over 2\alpha}(\partial_\mu \mathscr{A}^\mu)^2
\nonumber\\&
+ i \bar{\mathscr{C}} \cdot \partial_\mu \mathscr{D}^\mu[\mathscr{A}]\mathscr{C}
+ {\alpha \over 8}g^2 (i \mathscr{C} \times \bar{\mathscr{C}}) \cdot (i \mathscr{C} \times \bar{\mathscr{C}}) .
\end{align}
\end{equation}
Although the NL field $\mathscr{B}$ can be removed by performing the Gaussian integration, we retain it to discuss the breaking of BRST symmetry, since 
$\langle \mathscr{B} \rangle$ could be an order parameter for the BRST symmetry breaking as shown shortly.
(Note that in order to perform the integration over the NL field, we must shift the $\mathscr{B}$ field.  If the $\mathscr{B}$ field has a non-vanishing vacuum expectation value (VEV), this procedure is nontrivial.)
  We define the potential $V(\mathscr{B})$ for the bosonic field $\mathscr{B}$ by
\begin{equation}
 V(\mathscr{B}) = -{\alpha \over 2}\left( \mathscr{B} - {1 \over 2}ig (\mathscr{C} \times \bar{\mathscr{C}}) + {1 \over \alpha} \partial_\mu \mathscr{A}^\mu \right)^2 .
\end{equation}
This potential has an absolute minimum (maximum) for negative (positive) value of $\alpha$ at
\begin{equation}
  \mathscr{B} = {1 \over 2}ig (\mathscr{C} \times \bar{\mathscr{C}}) - {1 \over \alpha} \partial_\mu \mathscr{A}^\mu  .
\end{equation}
By keeping the Lorentz invariance, we have
$\langle \mathscr{A}^\mu \rangle = 0$
and hence the vacuum state $|0 \rangle$ obeys
\begin{equation}
  \langle 0|\mathscr{B}^A|0 \rangle  = {1 \over 2}g \langle 0|i(\mathscr{C} \times \bar{\mathscr{C}})^A |0 \rangle  .
\end{equation}
\par
Suppose that the ghost--anti-ghost condensation occurs, 
\begin{equation}
   \langle 0|i(\mathscr{C} \times \bar{\mathscr{C}})^A |0 \rangle \not= 0 .
\end{equation}
Then the NL field $\mathscr{B}$ acquires the non-vanishing VEV and the spontaneous breaking of the BRST symmetry occurs, since
\begin{equation}
  \langle 0| \bm{\delta}_B \bar{\mathscr{C}}^A |0 \rangle := 
\langle 0| \{ i Q_B, \bar{\mathscr{C}}^A \} |0 \rangle 
\equiv i \langle 0| \mathscr{B}^A |0 \rangle 
= {1 \over 2}i g \langle 0|i(\mathscr{C} \times \bar{\mathscr{C}})^A |0 \rangle  
\not= 0 .
\label{QB1}
\end{equation}
In this case, the anti-BRST symmetry is also spontaneously broken simultaneously, since
\begin{equation}
  \langle 0| \bar{\bm{\delta}}_B \mathscr{C}^A |0 \rangle := 
\langle 0| \{ i \bar{Q}_B, \mathscr{C}^A \} |0 \rangle 
\equiv i \langle 0| \bar{\mathscr{B}}^A |0 \rangle
= {1 \over 2}i g \langle 0|i(\mathscr{C} \times \bar{\mathscr{C}})^A |0 \rangle  ,
\label{QB2}
\end{equation}
where we have used
\begin{equation}
  \langle 0| \bar{\mathscr{B}} |0 \rangle = -\langle 0| \mathscr{B} |0 \rangle + ig \langle 0| (\mathscr{C} \times \bar{\mathscr{C}}) |0 \rangle .
\end{equation}
The Noether current for the BRST symmetry is obtained as
$J_B^\mu = - \mathscr{F}^{\mu\nu} \cdot \bm{\delta}_B \mathscr{A}_\nu 
+ \mathscr{B} \cdot \bm{\delta}_B \mathscr{A}^\mu
+ i \partial^\mu \bar{\mathscr{C}} \cdot \bm{\delta}_B \mathscr{C}$.
By making use of the equal-time canonical commutation relation for the anti-ghost, $
 \{ i(\mathscr{D}_0 \mathscr{C})^A(\bm{x},t), \bar{C}^B(\bm{y},t) \} = i \delta^{AB} \delta^3(\bm{x}-\bm{y}) 
$, the relation (\ref{QB1}) is recovered as 
\begin{equation}
 \langle 0| \{ iQ_B, \bar{\mathscr{C}}^B(\bm{y},t) \}|0 \rangle
 \equiv  \int d^3\bm{x} \langle 0| \{ i J_B^0(\bm{x},t), \bar{\mathscr{C}}^B(\bm{y},t) \}|0 \rangle
 = i  \langle 0| \mathscr{B}^A(\bm{y},t)  |0 \rangle .
\end{equation}
The similar expression is also obtained for the anti-BRST symmetry in agreement with (\ref{QB2}).

\par
In order to determine whether such a ghost--anti-ghost condensation and the resulting spontaneous breakdown of BRST supersymmetry take place or not, it is important to evaluate the effective potential for the composite operator 
$i(\mathscr{C} \times \bar{\mathscr{C}})^A $.  For $G=SU(2)$, the similar analysis to that performed in the paper~\cite{KS00} leads to
the total bosonic effective potential:
\begin{equation}
V(\mathscr{B},\varphi)=V(\varphi)
-{\alpha \over 2}\left( \mathscr{B} + {1 \over 2\alpha g} \varphi  \right)^2 ,
\end{equation}
where the effective potential $V(\varphi)$ of 
$\varphi^A \sim -\alpha g^2 \langle 0|i(\mathscr{C} \times \bar{\mathscr{C}})^A |0 \rangle$ 
is given by
\begin{equation}
 V(\varphi) = {1 \over \alpha g^2} |\varphi|^2 
 + {1 \over 32\pi^2} |\varphi|^2 \left[ \ln \left( {|\varphi| \over 4\pi \mu^2}  \right)^2 + C\right] ,
\end{equation}
with a regularization-dependent constant $C$ and the renormalization scale $\mu$. 
It is easy to see that the potential $V(\varphi)$ has stationary points 
for
$\mathscr{B}^A = - {1 \over 2\alpha g} \varphi^A$
and
$|\varphi|:= \sqrt{\varphi^A \varphi^A} = \varphi_0$
where
$
\varphi_0= 4\pi \mu^2 e^{-(1+C)/2} \exp \left[ -{8\pi^2 \over \alpha g^2(\mu)} \right]$.  Note that $\varphi_0$ exhibits the pathological behavior for $\alpha<0$; $\varphi_0 \uparrow \infty$ as $g \downarrow 0$.
Hence the ghost--anti-ghost condensation is meaningful only when $\alpha>0$. 
However, the total bosonic effective potential $V(\mathscr{B},\varphi)$ does not have any absolute minimum for $\alpha>0$. Therefore, we can not conclude that either the BRST symmetry or the anti-BRST symmetry is spontaneously broken (Here note that we don't exclude the symmetry breaking due to other mechanisms). 
\par
In the case of $\alpha=0$, this gauge is reduced to the Landau gauge in the Lorentz gauge fixing.
In the gauge fixing of the Lorentz type, the GF+FP term is given by
\begin{equation}
\begin{align}
 \mathscr{L}_{GF+FP} &= - i \bm{\delta}_B \left( \bar{\mathscr{C}} \cdot \left( \partial_\mu \mathscr{A}^\mu + {\alpha \over 2}\mathscr{B}  \right) \right) 
\nonumber\\
 &= \mathscr{B} \cdot \partial_\mu \mathscr{A}^\mu + i \bar{\mathscr{C}} \cdot \partial_\mu \mathscr{D}^\mu[\mathscr{A}]\mathscr{C}
+ {\alpha \over 2} \mathscr{B} \cdot \mathscr{B} ,
\end{align}
\end{equation}
which reads
\begin{equation}
 \mathscr{L}_{GF+FP} = 
   {\alpha \over 2} \left( \mathscr{B} + {1 \over \alpha} \partial_\mu \mathscr{A}^\mu \right)^2
- {1 \over 2\alpha}(\partial_\mu \mathscr{A}^\mu)^2
+ i \bar{\mathscr{C}} \cdot \partial_\mu \mathscr{D}^\mu[\mathscr{A}] \mathscr{C} .
\end{equation}
The minimum or maximum of the potential of the NL field $\mathscr{B}$ occurs at $\mathscr{B}=0$, which implies $\langle 0| \mathscr{B}^A |0 \rangle = 0$. Therefore, in the Lorentz gauge the spontaneous BRST symmetry breaking due to the mechanism mentioned above can not occur (at least in the tree level).
The above result guarantees the conventional treatment of making use of the BRST symmetry in the Lorentz gauge.  (up to the Gribov problem discussed in the final section).

\par
If a continuous symmetry is spontaneously broken,
there appears a massless particle called the Nambu-Goldstone particle, according to the Nambu-Goldstone theorem.
Consequently, the ghost and anti-ghost could be identified as the Nambu-Goldstone particle and hence they should be massless. 
Although the above consideration cannot yield any result on the gluon mass, gluons are expected to be massive in this gauge at least in the pure gauge sector~\cite{HK85}.  This situation is in sharp contrast with the modified MA gauge in which the off-diagonal ghost as well as off-diagonal gluons become massive  (in consistent with the infrared Abelian dominance) as argued in the paper~\cite{KS00}.
\par

\section{\label{sec:mMAgauge}Modified MA gauge}

\par
First of all, we decompose the gauge potential into the diagonal and off-diagonal components,
$
  \mathscr{A}_\mu(x) = \mathscr{A}_\mu^A(x) T^A = a_\mu^i(x) T^i + A_\mu^a(x) T^a ,
$
where $T^i \in \mathscr{H}$ and $T^a \in \mathscr{G}-\mathscr{H}$ with $\mathscr{H}$ being the Cartan subalgebra of the Lie algebra $\mathscr{G}$.
For gauge fixing of off-diagonal components, we adopt the  maximal Abelian (MA) gauge.
The MA gauge is a nonlinear gauge, whereas the Lorentz gauge is a linear gauge.  It is known~\cite{MLP85} that the introduction of the quartic ghost interaction is necessary to preserve the renormalizability of the Yang-Mills theory, since such an interaction is generated through radiative corrections even if the original Lagrangian does not involve such a term.  In fact, the quartic ghost interaction can be introduced as a BRST exact form:
\begin{equation}
  \mathscr{L}_{GF+FP}' = -  i \bm{\delta}_B \left[ 
\bar C^a  \left\{ D_\mu[a]A^\mu + {\alpha \over 2} B \right\}^a
- i {\zeta \over 2} g f^{abi} \bar C^a \bar C^b  C^i
- i {\zeta \over 4} g f^{abc} C^a \bar C^b \bar C^c \right] .
\label{GF2}
\end{equation}
By performing the BRST transformation explicitly, we obtain
\begin{equation}
\begin{align}
  \mathscr{L}_{GF+FP}' &=     
B^a D_\mu[a]^{ab}A^\mu{}^b+ {\alpha \over 2} B^a B^a
\nonumber\\
&+ i \bar C^a D_\mu[a]^{ac} D^\mu[a]^{cb} C^b
- i g^2 f^{adi} f^{cbi} \bar C^a C^b A^\mu{}^c A_\mu^d 
\nonumber\\
&+ i \bar C^a D_\mu[a]^{ac}(g f^{cdb}  A^\mu{}^d C^b)
+ i \bar C^a g  f^{abi} (D^\mu[a]^{bc}A_\mu^c) C^i 
\nonumber\\
&+{\zeta \over 8} g^2 f^{abe}f^{cde} \bar C^a \bar C^b C^c C^d
+ {\zeta \over 4} g^2 f^{abc} f^{aid} \bar C^b \bar C^c C^i C^d
+ {\zeta \over 2} g f^{abc} i B^b C^a \bar C^c  
\nonumber\\
&- \zeta  g f^{abi} i B^a \bar C^b C^i 
+ {\zeta \over 4} g^2 f^{abi} f^{cdi} \bar C^a \bar C^b C^c C^d \} .
\label{GF3}
\end{align}
\end{equation}
In particular, the $SU(2)$ case is greatly simplified as ($a,b,c,d=1,2$)
\begin{equation}
\begin{align}
  \mathscr{L}_{GF+FP}' &=   
B^a D_\mu[a]^{ab}A^\mu{}^b+ {\alpha \over 2} B^a B^a
\nonumber\\
&+ i \bar C^a D_\mu[a]^{ac} D^\mu[a]^{cb} C^b
- i g^2 \epsilon^{ad} \epsilon^{cb} \bar C^a C^b A^\mu{}^c A_\mu^d 
\nonumber\\
&
+ i \bar C^a g  \epsilon^{ab} (D_\mu[a]^{bc}A_\mu^c) C^3 
\nonumber\\
&- \zeta  g \epsilon^{ab} i B^a \bar C^b C^3 
+ {\zeta \over 4} g^2 \epsilon^{ab} \epsilon^{cd} \bar C^a \bar C^b C^c C^d .
\label{GF4}
\end{align}
\end{equation}
However, the strength $\zeta$ of the quartic ghost self-interaction
${\zeta \over 4} g^2 f^{abi} f^{cdi} \bar C^a \bar C^b C^c C^d$ is arbitrary.  
Now we adopt the modified version of the MA gauge proposed by the author~\cite{KondoII},
\begin{equation}
 \mathscr{L}_{GF+FP} = i \bm{\delta}_B \bar{\bm{\delta}}_B \left[ {1 \over 2}A_\mu^a(x) A^{\mu}{}^a(x)
 - {\alpha \over 2}i C^a(x) \bar{C}^a(x) \right] ,
\label{MAGF}
\end{equation}
where $\alpha$ corresponds to the gauge fixing parameter for the off-diagonal components, since the explicit calculation of the anti-BRST transformation $\bar{\bm{\delta}}_B$ yields
\begin{equation}
  \mathscr{L}_{GF+FP} = -  i \bm{\delta}_B \left[ 
\bar C^a  \left\{ D_\mu[a]A^\mu + {\alpha \over 2} B \right\}^a
- i {\alpha \over 2} g f^{abi} \bar C^a \bar C^b  C^i
- i {\alpha \over 4} g f^{abc} C^a \bar C^b \bar C^c \right] .
\label{GF20}
\end{equation}
(The most general form of the MA gauge has been obtained by Hata and Niigata~\cite{HN93}.)
In the modified MA gauge, the requirement of the orthosymplectic ($OSp(4|2)$) invariance yields the quartic ghost interaction and simultaneously determines the strength $\zeta$ as $\zeta=\alpha$.
For $G=SU(2)$, the modified MA gauge is further rewritten as
\begin{equation}
\begin{align}
  \mathscr{L}_{GF+FP} &=   
  {\alpha \over 2} \left( B^a -  g \epsilon^{ab} i \bar C^b C^3 
+ {1 \over \alpha}D_\mu^{ab}[a]A^\mu{}^b \right)^2
-{1 \over 2\alpha}(D_\mu[a]^{ab}A^\mu{}^b)^2  
\nonumber\\
& 
+ i \bar C^a D_\mu[a]^{ac} D^\mu[a]^{cb} C^b
- i g^2 \epsilon^{ad} \epsilon^{cb} \bar C^a C^b A^\mu{}^c A_\mu^d 
\nonumber\\
& 
+ {\alpha \over 4} g^2 \epsilon^{ab} \epsilon^{cd} \bar C^a \bar C^b C^c C^d .
\label{GF6}
\end{align}
\end{equation}

In order to completely fix the gauge degrees of freedom, we must add the GF+FP term for the diagonal component $a_\mu^i$ to (\ref{GF20}).
If we  adopt the gauge fixing condition of the Lorentz type, $\partial^\mu a_\mu^i=0$ for the diagonal components,%
\footnote{Even for $G=U(1)^n$, the GF+FP term is written in the manifest $OSp(4|2)$ invariant form~\cite{KondoIII}
\begin{equation}
  \mathscr{L}_{GF+FP}^{Abelian} =  i \bm{\delta}_B \bar{\bm{\delta}}_B \left[ 
{1 \over 2}a_\mu^i a^\mu{}^i - {\beta \over 2}i c^i \bar c^i
 \right] 
= - a_\mu^i \partial^\mu B^i + {\beta \over 2}B^i B^i 
- i \partial_\mu \bar c^i \partial^\mu c^i .
\label{GFabel2}
\end{equation}
}
the GF+FP term reads
\begin{equation}
  \mathscr{L}_{GF+FP}^{diag} = -  i \bm{\delta}_B \left[ 
\bar c^i  \left( \partial^\mu a_\mu^i + {\beta \over 2} B^i \right)
 \right] 
= - a_\mu^i \partial^\mu B^i +{\beta \over 2}B^i B^i -i \partial_\mu  \bar c^i (D^\mu[A] c)^i .
\label{GFabel}
\end{equation}
We can rewrite it as
\begin{equation}
\begin{align}
  \mathscr{L}_{GF+FP}^{diag} 
&= {\beta \over 2}\left( B^i+{1 \over \beta}\partial^\mu a_\mu \right)^2 
- {1 \over 2\beta}(\partial^\mu a_\mu)^2 
-i \partial_\mu  \bar c^i (D^\mu[A] c)^i 
\\
&= {\beta \over 2}\left( \bar B^i-ig(C\times \bar C)^i +{1 \over \beta}\partial^\mu a_\mu \right)^2 
- {1 \over 2\beta}(\partial^\mu a_\mu)^2 
-i \partial_\mu  \bar c^i (D^\mu[A] c)^i .
\label{GFabel3}
\end{align}
\end{equation}

\par
The potential for the off-diagonal NL field $B^a$ has  a minimum (maximum) for negative (positive) value of $\alpha$ at
\begin{equation}
  B^a =  g \epsilon^{ab} i \bar C^b C^3 
- {1 \over \alpha}D_\mu^{ab}[a]A^\mu{}^b  .
\end{equation}
The vacuum $|0 \rangle$ obeys
\begin{equation}
  \langle 0|B^a|0 \rangle  =  g \langle 0| \epsilon^{ab} i \bar C^b C^3 |0 \rangle - {1 \over \alpha} g \langle 0|\epsilon^{ab} a^\mu A_\mu^b  |0 \rangle .
\label{Bvev}
\end{equation}
The ghost--anti-ghost condensation 
$\varphi^a := \langle 0| \epsilon^{ab} i \bar C^b C^3 |0 \rangle \not= 0$ 
will not occur,%
\footnote{This condensation is not invariant under the residual $U(1)$ gauge symmetry.  Hence, it is excluded in the previous treatment in which the residual U(1) is preserved.}
 because such a condensation leads to a non-zero VEV of $B^a$ and anti-ghost $\bar C^a$ is identified with the massless Nambu-Goldstone particle according to
\begin{equation}
  \langle 0| \bm{\delta}_B \bar C^a |0 \rangle := 
\langle 0| \{ i Q_B, \bar C^a \} |0 \rangle 
\equiv i \langle 0| B^a |0 \rangle  .
\end{equation}
This contradicts with the result~\cite{KS00} that the off-diagonal ghosts become massive due to ghost--anti-ghost condensation.  
The second term in the right-hand-side of (\ref{Bvev}) is also zero, because it is not invariant under the residual U(1) symmetry.
Therefore, we conclude 
$\langle 0| B^a |0 \rangle = 0$.
\par
Suppose that the ghost--anti-ghost condensation occurs in the sense 
\begin{equation}
 \varphi^i := \langle 0| i(C \times \bar C)^i |0 \rangle 
\equiv \langle 0| if^{iab} C^a \bar C^b |0 \rangle \not= 0 .
\label{off-cond}
\end{equation}
Then the anti-BRST symmetry is spontaneously broken, since
\begin{equation}
\begin{align}
  \langle 0| \bar{\bm{\delta}}_B C^i |0 \rangle &:=
\langle 0| \{ i \bar Q_B, C^i \} |0 \rangle 
\equiv i \langle 0| \bar B^i |0 \rangle
\nonumber\\
&= -i \langle 0| B^i |0 \rangle 
+ ig \langle 0| i(C \times \bar C)^i |0 \rangle  
\not= 0,
\end{align}
\end{equation}
where we have assumed $ \langle 0| B^i |0 \rangle=0$.
\par
It should be remarked that we have excluded a possibility that the diagonal NL field $B^i$ acquires a non-zero VEV in such a way that the non-zero value of $\langle 0| B^i |0 \rangle$ exactly cancels the ghost--anti-ghost condensation, 
$g \langle 0| i(C \times \bar C)^i |0 \rangle$, leading to 
$\langle 0| \bar B^i |0 \rangle = 0$.
In this case, the BRST symmetry is spontaneously broken, whereas the anti-BRST symmetry remains unbroken.
In fact, an arbitrary non-zero VEV,
$\langle 0| B^i |0 \rangle \not=0$ leads to the spontaneous breakdown of the BRST symmetry as
\begin{equation}
  \langle 0| \bm{\delta}_B \bar C^i |0 \rangle := 
\langle 0| \{ i Q_B, \bar C^i \} |0 \rangle 
= i \langle 0| B^i |0 \rangle   
\not= 0 .
\end{equation}
This argument suggests that the BRST symmetry or the anti-BRST symmetry is spontaneously broken, if the condensation occurs,
$\varphi^i=\langle 0| if^{iab} C^a \bar C^b |0 \rangle \not= 0$.
\par
Here we recall that the BRST charge $Q_B$ is a generator of the translation in the Grassmann coordinate $\theta$, while the anti-BRST charge $\bar Q_B$ is a generator of the translation in $\bar \theta$ in the superspace $(x^\mu, \theta, \bar \theta)$ as shown by Bonora and Tonin~\cite{BT81}.
The modified MA gauge fixing deals with the FP ghost and anti-ghost on equal footing.  Therefore, it seems unnatural that only the anti-BRST symmetry is spontaneously broken, while the BRST symmetry remains unbroken and vice versa.
The modified MA gauge has a hidden orthosymplectic symmetry $OSp(4|2)$, if $\alpha\not=0$ as demonstrated in~\cite{KondoII}.
Actually, if the orthosymplectic symmetry $OSp(4|2)$ is not broken, we can rotate $\theta$ to $\bar \theta$ and vice versa.
This argument suggest that both BRST and anti-BRST symmetry can be spontaneously broken, if the ghost--anti-ghost condensation of the type
$\langle 0| if^{iab} C^a \bar C^b |0 \rangle \not= 0$
takes place.%
\footnote{ 
By using the Noether current $J_B^\mu$ for the BRST symmetry  and the equal-time canonical commutation relation for the anti-ghost, i.e.,
$
 \{ \pi_{\bar{C}}^i(\bm{x},t), \bar{C}^i(\bm{y},t) \} = i \delta^{ij} \delta^3(\bm{x}-\bm{y}) ,
$
it is possible to show that  
$
 \langle 0| \{ iQ_B, \bar{C}^i(\bm{y},t) \}|0 \rangle
 \equiv  \int d^3\bm{x} \langle 0| \{ i J_B^0(\bm{x},t), \bar{C}^i(\bm{y},t) \}|0 \rangle
 = i \langle 0| B^i |0 \rangle .
$
Here we have used the explicit form,
$
 J_B^\mu = - \mathcal{F}^{\mu\nu} \cdot \bm{\delta}_B \mathscr{A}_\nu
 + \mathscr{B} \cdot \bm{\delta}_B \mathscr{A}^\mu 
 + i \partial^\mu \bar{\mathscr{C}} \cdot \bm{\delta}_B \mathscr{C} 
 - ig \epsilon^{ab} \bar{C}^a (\bm{\delta}_B A^\mu{}^b)C^3 
 - ig \epsilon^{ab} \bar{C}^a (\bm{\delta}_B a^\mu) C^b
 - 2ig \epsilon^{ab} \bar{C}^a a^\mu \bm{\delta}_B C^b .
$
The similar result holds also for the anti-BRST symmetry. 
}
\par
An advantage of the spontaneous BRST supersymmetry breaking is as follows.
The Nambu-Goldstone particle (fermion) associated with the spontaneous breaking of the BRST $Q_B$ or anti-BRST $\bar Q_B$ symmetry is identified with the diagonal anti-ghost $\bar C^i$ or diagonal ghost $C^i$, respectively.  Hence the diagonal ghost and anti-ghost are massless.
This is consistent with the infrared Abelian dominance which is expected to be realized, if the off-diagonal components of gluons and ghosts become massive while the diagonal components remain massless.
\par
A disadvantage is as follows.
If both BRST and anti-BRST symmetry are spontaneously broken, 
the vacuum is not invariant under the BRST symmetry $Q_B|0 \rangle \not= 0$, $\bar Q_B|0 \rangle \not= 0$, and
we can not conclude the gauge parameter $\alpha$ independence of the vacuum-to-vacuum amplitude and the physical $S$ matrix\cite{MLP85} (i.e., ${\delta Z \over \partial \alpha}\not=0$), since 
\begin{equation}
\begin{align}
 {\delta Z \over \partial \alpha}
&= {\delta  \over \partial \alpha} \langle 0; out| 0; in \rangle
\nonumber\\
&= i{1 \over 2} \int d^4x \langle 0; out| \bm{\delta}_B \bar{\bm{\delta}}_B \left(  C^a(x) \bar C^a(x) \right) |0; in \rangle
\nonumber\\
&= i{1 \over 2} \int d^4x \langle 0; out| \{ Q_B, [ \bar Q_B,   C^a(x) \bar C^a(x)   ] \} |0; in \rangle
\nonumber\\
&= - i{1 \over 2} \int d^4x \langle 0; out| \{ \bar Q_B, [ Q_B, C^a(x) \bar C^a(x)  ] \} |0; in \rangle .
\end{align}
\end{equation}
Similarly, we find the gauge parameter $\beta$ dependence as
\begin{equation}
\begin{align}
 {\delta Z \over \partial \beta}
&= {\delta  \over \partial \beta} \langle 0; out| 0; in \rangle
\nonumber\\
&=  {1 \over 2} \int d^4x \langle 0; out| \bm{\delta}_B \left(  \bar C^i(x) B^i(x) \right) |0; in \rangle
\nonumber\\
&= {1 \over 2} \int d^4x \langle 0; out| \{ i Q_B, \bar C^i(x) B^i(x) \} |0; in \rangle .
\end{align}
\end{equation}
Therefore, the Yang-Mills theories with different gauge parameters are different theories.  The parameter $\alpha$ should be determined based on other arguments.  As a way to determine the value of $\alpha$, $\alpha$ could be chosen to sit on the fixed point of the renormalization group as proposed in~\cite{KS00}.

\par 
Now we examine a simple case $G=SU(2)$.
The total bosonic effective potential is given by
\begin{equation}
 V(\bar{B}, B^a, \varphi) = V(\varphi) - \frac{\beta}{2}\left( \bar{B} + {1 \over \zeta g} \varphi \right)^2 - \frac{\alpha}{2} B^a B^a  ,
\end{equation}
where we have ommitted the index of the diagonal component, i.e., $\bar{B}\equiv \bar{B}^3$, $\varphi \equiv \varphi^3$
and $V(\varphi)$ is given by\cite{KS00}
\begin{equation}
 V(\varphi) = {1 \over 2\zeta g^2} \varphi^2 
 + {1 \over 32\pi^2} \varphi^2 \left[ \ln \left( {1 \over 4\pi \mu} \varphi \right)^2 + \text{const.} \right] .
\end{equation}
Note that $V(\varphi)$ has  minima at non-zero values of $\varphi$, $\varphi=\pm \varphi_0$, for $\zeta>0$.  Therefore, the spontaneous breakdown of the BRST or anti-BRST symmetry could happen, if $\zeta>0$ and $\beta<0$, since the total bosonic effective potential has an absolute minimum at non-zero value of $\bar{B}=-(\zeta g)^{-1}\varphi$.
A simple way to avoid this situation is to take $\beta>0$.
If $\beta>0$, even the non-zero condensation $\varphi \not= 0$ for $\zeta>0$ does not leads to the spontaneous breakdown of the BRST symmetry, since it corresponds to the saddle point.
However, $\zeta>0$ and $\alpha<0$ can not be simultaneously realized if we impose the $OSp(4|2)$ invariance which postulates $\zeta=\alpha$.  Thus the modified MA gauge with $OSp(4|2)$ invariance does not cause the spontaneous breakdown of the BRST and the anti-BRST supersymemtry, at least based on the mechanism of ghost and anti-ghost condensation discussed above. 
\par
On the other hand, another type of ghost--anti-ghost condensation 
\begin{equation}
 \langle 0| (C \cdot \bar C)_{off}|0 \rangle 
:= \langle 0| \delta^{ab} C^a \bar C^b |0 \rangle \not= 0 ,
\label{diag-cond}
\end{equation}
is possible to occur.  However, it does not lead to the spontaneous breaking of BRST or anti-BRST symmetries.  Therefore, if the situation
\begin{equation}
  \langle 0| \delta^{ab} C^a \bar C^b |0 \rangle \not= 0 ,
\quad
   \langle 0| f^{iab}  C^a \bar C^b |0 \rangle  = 0 ,
\end{equation}
is realized, we don't have any direct argument suggesting the breaking of BRST and/or anti-BRST symmetries. 
The diagonal condensation, 
$\langle 0| \delta^{ab} C^a \bar C^b |0 \rangle \not= 0$, 
is enough  for off-diagonal gluons and off-diagonal ghosts (and anti-ghost) to acquire their masses as shown in~\cite{KS00}.
However, we lose a chance of explaining the ghost--anti-ghost condensation as a spontaneous breaking of a global symmetry $SL(2,R)$~\cite{Schaden99} which is shown to exist at least for the $SU(2)$ Yang-Mills theory in the modified MA gauge.
It is a dynamical problem to determine which case is actually realized in the Yang-Mills theory in the modified MA gauge.  More systematic investigation will be given elsewhere.
\par
A basic ingredient of the above argument is the non-vanishing VEV of the composite operator $f^{iab} C^a(x) \bar C^b(x)$, i.e., 
$\varphi^i :=\langle 0| if^{iab} C^a \bar C^b |0 \rangle \not= 0$.
Note that the composite operator is not BRST invariant, 
$\bm{\delta}_B(f^{iab} C^a \bar C^b) \not= 0$.  However, this does not immediately mean the breakdown of BRST symmetry.
The BRST transformation does transform a field into a different field or a composite field.  Therefore, the non-zero $\varphi$ does not necessarily mean the breaking of the BRST symmetry $Q_B$.  This situation is quite similar to the Higgs field in the Higgs-Kibble model $\phi(x)$ (The Higgs field is not BRST invariant.  Nevertheless, it can have a non-vanishing VEV $\langle \phi \rangle$ without breaking the BRST symmetry after the gauge fixing).  This should be compared with the spin or scalar field models with global rotational symmetry (If the scalar has a non-vanishing VEV in a direction, the global rotational symmetry of the original Lagrangian is spontaneously broken). 
Finally, we consider an exceptional case in which the composite operator is written as a BRST transform of an operator $\mathcal{O}$, i.e.,
$f^{iab} C^a \bar C^b = \bm{\delta}_B \mathcal{O}$.  If this is the case, the non-zero VEV leads to the spontaneous breakdown of the BRST symmetry.  However, this situation is not realized for this operator.  The reason is as follows.  The composite operator written in the BRST exact form  is BRST invariant, 
$
 \bm{\delta}_B(f^{iab} C^a \bar C^b) 
= \bm{\delta}_B \bm{\delta}_B \mathcal{O}=0
$ due to nilpotency of BRST transformation.  This result contradicts with the non BRST-invariance of the composite operator.
Thus there does not exist an operator $\mathcal{O}$ such that $f^{iab} C^a \bar C^b = \bm{\delta}_B \mathcal{O}$.  Thus this case is excluded.

\section{\label{sec:Gribov}Comments on the Gribov problem}

In the rest of this paper, we discuss an implication of the above result to the Gribov problem.
The models treated in this paper include quartic ghost self-interaction terms whose coefficients (strengths) are proportional to $\alpha$.  For non-zero $\alpha$, therefore, the integration of the ghost and anti-ghost field does not necessarily yield the conventional Faddeev-Popov determinant.  For these models, thus, we can not find the exact relationship between the spontaneous breaking of BRST supersymmetry and the Gribov problem (i.e., existence of the zero eigenvalue of the operator in the Faddeev-Popov determinant), contrary to the conventional Lorentz type gauge which has been extensively discussed by Fujikawa~\cite{Fujikawa83}.
However, this does not mean that the inclusion of quartic ghost interaction can avoid the Gribov problem.  The quartic ghost interaction can be reduced to a bilinear form in the ghost and anit-ghost field by introducing the auxiliary field $\varphi$.  Then we encounter the Gribov problem even in the modified MA gauge, see~\cite{BHWT00} for details.
\par
In view of this, we recall the Witten index~\cite{Witten82}.
It is known that the vanishing of the Witten index,
${\rm Tr}[(-1)^F]={\rm Tr}[\exp(i\pi Q_c)]=0$,
is a necessary condition for spontaneous supersymmetry breaking (although it is not a sufficient condition).
The regularized index
\begin{equation}
 Z(\beta) := {\rm Tr}(e^{-\beta H+i\pi Q_c}) 
\equiv {\rm STr}(e^{- \beta H}) 
\end{equation}
is written in the Euclidean path integral as~\cite{Fujikawa82,CG82,Nicolai81}
\begin{equation}
 Z(\beta) := {\cal N}^{-1} \int \mathscr{D}\mathscr{A}_\mu \mathscr{D}\mathscr{B} \mathscr{D}\mathscr{C} \mathscr{D}\bar{\mathscr{C}} 
\exp \left\{ \int_0^\beta d\tau \int d^3x \mathscr{L}_{tot}[\mathscr{A},\mathscr{B},\mathscr{C},\bar{\mathscr{C}}] \right\}  ,
\end{equation}
where periodic boundary conditions must be imposed on {\it all} the fields in Euclidean time $\tau$ to avoid the explicit breaking of the BRST supersymmetry (as emphasized by Fujikawa).
After the BRST supersymmetry is spontaneously broken, we have more fermion than bosons with zero eigenvalues, then 
$Z(\beta)=0$.    
Before the spontaneous breaking, we have exactly degenerate eigen values for fermions and bosons and thus $Z(\beta)=1$.
Thus, the spontaneous breakdown of BRST supersymmetry leads to the (extra) zero eigenvalue. 
\par
If the Witten index is non-zero, the BRST symmetry is not spontaneously broken.  As a result, the off-diagonal ghost condensation (\ref{off-cond}) is prohibited and the diagonal ghost condensation (\ref{diag-cond}) can be allowed to be non-zero, which is necessary for the off-diagonal ghost and gluons to acquire non-zero masses.  
In view of these, it is desirable to calculate the Witten index~\cite{Witten82} precisely.

\section*{Acknowledgments}
The author would like to thank Kazuo Fujikawa and Shogo Tanimura for helpful conversations on the BRST symmetry breaking and the Gribov problem in the early stage of the investigation.
This work is supported in part by
a Grant-in-Aid for Scientific Research from the Ministry of
Education, Science and Culture (10640249).



\baselineskip 14pt

\begin{thebibliography}{99}

\bibitem{BRST}
  C. Becchi, A. Rouet and R. Stora,
Renormalization of the Abelian Higgs model,
  Commun. Math. Phys. 42, 127-162 (1975);
Renormalization of gauge theories,
  Ann. Phys. 98, 287-321 (1976).
\\
I.V. Tyutin,
  Lebedev preprint, FIAN No.39 (in Russian) (1975).

\bibitem{KO79}
T.~Kugo and I.~Ojima,
Manifestly covariant canonical formulation of Yang-Mills theories physical state subsidiary conditions and physical S-matrix unitarity,
Phys. Lett. B 73, 459-462 (1978).
Local covariant operator formalism of nonabelian gauge theories And quark confinement problem,
Prog.\ Theor.\ Phys.\ Suppl.\ {\bf 66}, 1-130 (1979).

\bibitem{antiBRST}
  G. Curci and R. Ferrari,
  Slavnov transformations and supersymmetry,
  Phys. Lett. B 63, 91-94 (1976).
  \\
I. Ojima,
  Another BRS transformation,
  Prog. Theor. Phys. 64, 625-638 (1980).

\bibitem{BT81}
  L. Bonora and M. Tonin,
Superfield formulation of extended BRS symmetry,
  Phys. Lett. B 98, 48-50 (1981).

\bibitem{Fujikawa83}
  K. Fujikawa,
Dynamical stability of the BRS supersymmetry and the Gribov problem,
Nucl. Phys. B 223, 218-234 (1983).


\bibitem{DJ82}
  R. Delbourgo and P.D. Jarvis,
Extended BRS invariance and OSP(4/2) supersymmetry,
J. Phys. A: Math. Gen. 15, 611-625 (1982).

\bibitem{BT82}
  L. Baulieu and J. Thierry-Mieg,
The principle of BRS symmetry: An alternative approach to Yang-Mills theories,
Nucl. Phys. B 197, 477-508 (1982).

\bibitem{Baulieu85}
L. Baulieu,
Perturbative gauge theories,
Phys. Reports 129, 1-74 (1985).

\bibitem{HK85}
  H. Hata and T. Kugo,
Color confinement, Becchi--Rouet--Stora symmetry, and negative dimensions,
Phys. Rev. D32, 938-944 (1985).







\bibitem{KondoI}
  K.-I. Kondo,
  Abelian-projected effective gauge theory of QCD with asymptotic
freedom and quark confinement,
  hep-th/9709109,
  Phys. Rev. D 57, 7467-7487 (1998).
  \\
K.-I. Kondo, 
  hep-th/9803063,
  Prog. Theor. Phys. Supplement, No. 131, 243-255.


\bibitem{KondoII}
  K.-I. Kondo, 
  Yang-Mills theory as a deformation of topological field
theory, dimensional reduction and quark confinement,
  hep-th/9801024,
  Phys. Rev. D 58, 105019 (1998).

\bibitem{KondoIII}
  K.-I. Kondo,
  Existence of confinement phase in quantum electrodynamics,
  hep-th/9803133,
  Phys. Rev. D 58, 085013 (1998).

\bibitem{KondoIV}
  K.-I. Kondo,
  Abelian magnetic monopole dominance in quark confinement,
  hep-th/9805153,
  Phys. Rev. D 58, 105016 (1998).

\bibitem{KondoV}
  K.-I. Kondo,
  Quark confinement and deconfinement in QCD from the viewpoint of
Abelian-projected effective gauge theory,
  hep-th/9810167,
  Phys. Lett. B 455, 251-258 (1999).

\bibitem{KondoVI}
  K.-I. Kondo,
  A formulation of the Yang-Mills theory as deformation of a topological field 
  hep-th/9904045,
  Intern. J. Mod. Phys. A (2001), in press.

\bibitem{Kondo00}
  K.-I. Kondo,
  Dual superconductivity, monopole condensation and confining string in low-energy Yang-Mills theory,
CHIBA-EP-123,
hep-th/0009152 (revised version in preparation).


\bibitem{MLP85}
  H. Min, T. Lee and P.Y. Pac,
  Renormalization of Yang-Mills theory in the abelian gauge,
  Phys. Rev. D 32, 440-449 (1985).



\bibitem{Schaden99}
  M. Schaden,
  Mass generation in continuum SU(2) gauge theory in covariant Abelian gauges,
  hep-th/9909011, 3rd revised version.

\bibitem{KS00}
  K.-I. Kondo and T. Shinohara,
  Abelian dominance in low-energy Gluodynamics due to dynamical mass generation,
  hep-th/0004158,
  Phys. Lett. B 491, 263-274 (2000).


\bibitem{tHooft81}
  G. 't Hooft,
  Topology of the gauge condition and new confinement
phases in non-Abelian gauge theories,
  Nucl.Phys. B 190 [FS3], 455-478 (1981).


\bibitem{dualsuper}
  Y. Nambu,
  Strings, monopoles, and gauge fields,
  Phys. Rev. D 10, 4262-4268 (1974).
\\
G. 't Hooft,
  in: High Energy Physics, edited by A. Zichichi 
(Editorice Compositori, Bologna, 1975).
\\
S. Mandelstam,
  Vortices and quark confinement in non-abelian gauge theories, 
 Phys. Report  23, 245-249 (1976).
\\
A.M. Polyakov,
  Compact gauge fields and the infrared catastrophe,
  Phys. Lett. B 59, 82-84 (1975).
  Quark confinement and topology of gauge theories,
  Nucl. Phys. B 120, 429-458 (1977).


\bibitem{Gribov78}
  V. Gribov,
Quantization of non-Abelian gauge theories,
Nucl. Phys. B 139, 1-19 (1978).


\bibitem{HN93}
  H. Hata and I. Niigata,
  Color confinement, abelian gauge and renormalization group flow,
  Nucl. Phys. B 389, 133-152 (1993).


\bibitem{BHWT00}
  F. Bruckmann, T. Heinzl, A. Wipf and T. Tok,
Instantons and Gribov copies in the maximally Abelian gauge,
hep-th/0001175,
Nucl. Phys. B 584, 589-614 (2000).

\bibitem{Witten82}
  E. Witten,
Constraints on supersymmetry breaking,
Nucl. Phys. B 202, 253-316 (1982).
Dynamical breaking of supersymmetry,
Nucl. Phys. B 185, 513-554 (1981).

\bibitem{Fujikawa82}
  K. Fujikawa,
Comment on the supersymmetry at finite temperature,
Z. Phys. C 15, 275 (1982).

\bibitem{CG82}
  S. Cecotti and L. Girardello,
Functional measure, topology and dynamical supersymmetry breaking,
Phys. Lett. B 110, 39-43 (1982).

\bibitem{Nicolai81}
  H. Nicolai,
A transformation formula for integrals over superfields,
Phys. Lett. B 101, 396-398 (1981).




\end{thebibliography}
\end{document}